\documentclass[letterpaper,english,reprint,aps]{revtex4-1}
\usepackage[T1]{fontenc}
\usepackage[latin9]{inputenc}
\usepackage{verbatim}
\usepackage{amsmath}
\usepackage{amssymb}

\makeatletter

\usepackage{graphicx}
\usepackage{xcolor}
\usepackage[bookmarks=false,linkcolor=blue,urlcolor=blue,colorlinks,citecolor=blue]{hyperref}
\usepackage{natbib}

\makeatother

\usepackage{babel}
\begin{document}

\title{Spontaneous Josephson $\pi$ Junctions with Topological Superconductors}

\author{Arbel Haim}

\affiliation{
Walter Burke Institute of Theoretical Physics, California Institute of Technology, Pasadena, CA 91125, USA\\
Department of Physics and the Institute for Quantum Information and Matter,\\
California Institute of Technology, Pasadena, CA 91125, USA
}

\date{\today}
\begin{abstract}
%We examine the question of whether a $\pi$ junction can spontaneously form in a Josephson junction between two topological superconductors.
We study a junction between two time-reversal-invariant
topological superconductors and show this system goes through a series
of multiple transitions between a $0$-junction phase, where the free
energy has its minimum for a superconducting phase difference of zero,
and a $\pi$-junction phase, where the free energy has its minimum
for a superconducting phase difference of $\pi$. These transitions
occur in the absence of Coulomb blockade or magnetic impurities. Rather, they are driven by the spin-orbit coupling in
the junction, and can be probed, for example, by measuring the tunneling
density of states or the critical current as a function of the junction's
length or its Fermi velocity. 
\end{abstract}
\maketitle

%\paragraph*{Introduction.\textemdash{}}
\section{Introduction}

Josephson $\pi$-junctions have been studied extensively in recent
decades~\citep{Cleuziou2006carbon,Kulik1966,Buzdin1982critical,Ryazanov2001coupling,Kontos2002Josephson,Schulz2000design,Franceschi2010hybrid,Zaikin2004some,Spivak1991negative}.
Unlike the more common Josephson $0$-junctions, where the free energy
is minimized by a phase difference of $\phi=0$, these are junctions
in which the free energy is minimized by a phase difference of $\phi=\pi.$

An important question is whether a $\pi$ junction can spontaneously form, in the absence of magnetic fields, in a  junction between two \emph{topological} superconductors. In the case of trivial superconductors, it has been established that a superconductor - quantum dot - superconductor
(S-QD-S) junction can exhibit $\pi$-junction behavior %when the 
as a result of Coulomb
interaction in the QD~\citep{Glazman1989resonant,Spivak1991negative,Yeyati1997resonant,Rozhkov2001josephson,Zaikin2004some,Siano2004Josephson,Choi2004Kondo,Schrade2018ParityWithComment}.

Recently, Josephson junctions with two time-reversal-invariant topological
superconductors (TRITOPSs) have also been studied~\citep{Zhang2013time,Liu2014non,Zhang2014anomalous,Kane2015the,Gong2016influence,Camjayi2017fractional,Arrachea2018catalogue}.
Such a topological superconductor~\citep{Schnyder2008classification,Qi2009time,Qi2010topological,Qi2011topological,Haim2019time}
hosts protected pairs of Majorana zero modes at each of its boundaries
while maintaining a bulk gap. It was shown that these Majorana zero
modes can form an effective spin which in turn screens the spin of the QD,
%can screen the Coulomb interaction in a S-QD-S junction
thereby avoiding the $\pi$-junction fate of conventional superconductors~\citep{Camjayi2017fractional}.

In this paper, we show that a Josephson junction with two TRITOPSs
can nevertheless be driven into the $\pi$-junction phase via a different
mechanism. Specifically, the system goes through multiple transitions
between 0-junction and $\pi$-junction behavior, as a function of
the rotation angle acquired by the electron's spin as it passes the
junction. This rotation is caused by spin-orbit coupling, and depends
on the junction's length and the Fermi velocity.

The mechanism behind the formation of the $\pi$ junction is intimately related to the defining topological property of the TRITOPS phase, namely the sign difference that exists between the pairing potentials of positive- and negative-helicity modes~\cite{Qi2010topological,Haim2019time}. As we show below, in the presence of spin-orbit coupling in the junction, this sign difference translates into a relative $\pi$ phase difference between the superconductors on the two sides of the junction (see Fig.~\ref{fig:phys_pic}).

We begin by studying a low-energy model, which provides a simple physical picture, and allows for an analytical expression describing the equilibrium phase difference. 
We then move on to study the junction numerically
using a microscopic lattice model. We use it to calculate the tunneling density
of states in the junction, and the critical current; these can serve
as experimental signatures of the $0$-$\pi$ transitions.

\begin{figure}
\begin{centering}
\begin{tabular}{lr}
\hskip -1.5mm
\includegraphics[clip=true,trim=0mm 1mm 0mm 0mm,width=4.3cm]{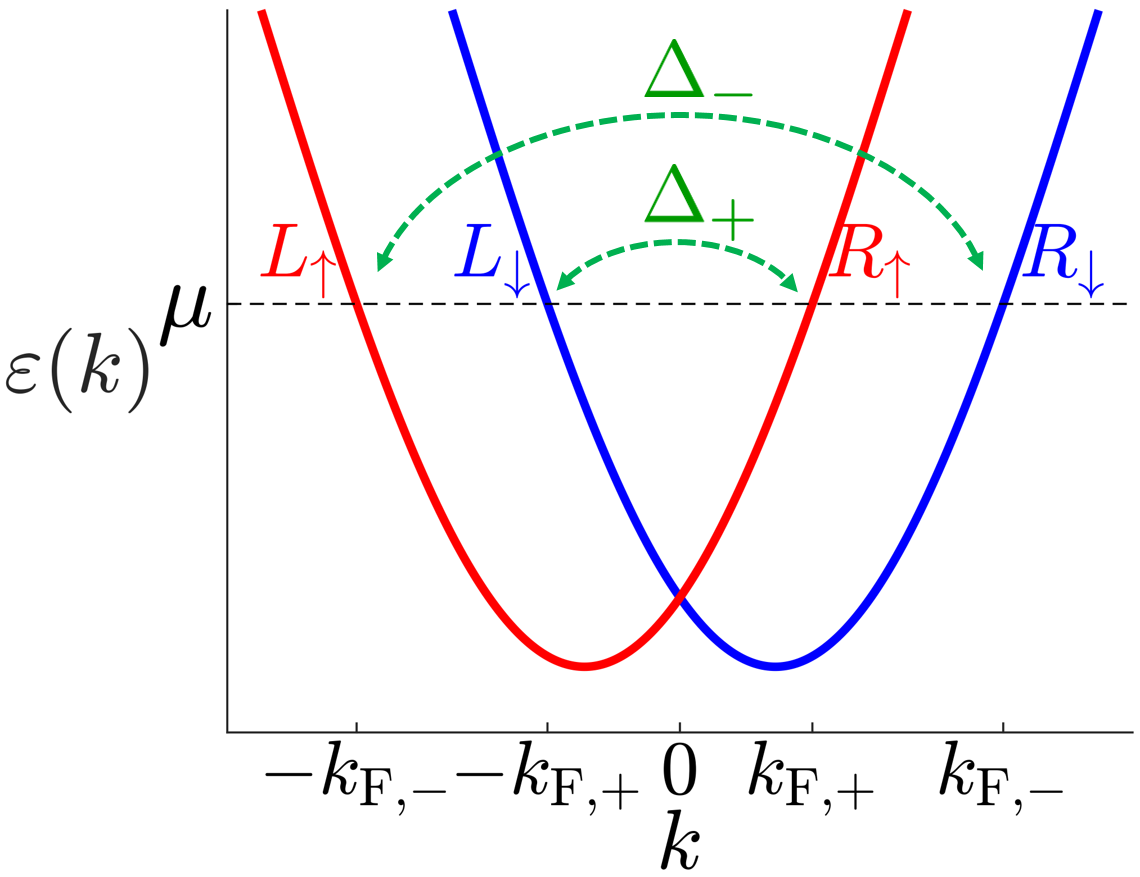}
\llap{\parbox[c]{8.2cm}{\vspace{-2mm}(a)}}
&
\includegraphics[clip=true,trim=0mm -3cm 0mm 0mm,width=4.25cm]{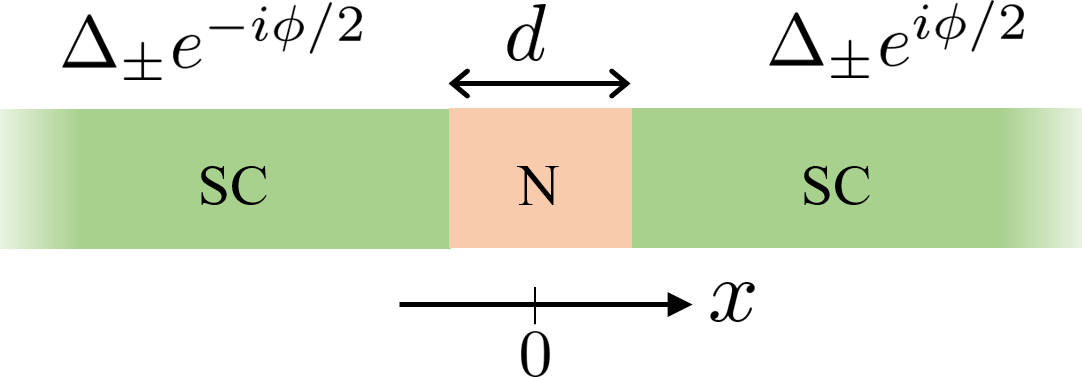}
\llap{\parbox[c]{8cm}{\vspace{-2mm}(b)}}
\end{tabular}
\end{centering}
\caption{(a) Electronic spectrum of the model described in Eq.~\eqref{eq:H_0_latt}. Close to the Fermi level, it can be described by four linearly-dispersing modes. The positive-helicity modes experience a pairing potential, $\Delta_+$, while the negative-helicity modes experience a pairing potential, $\Delta_-$ {[}see Eq.~(\ref{eq:H_0_low_E}){]}. The time-reversal-invariant topological
superconductor (TRITOPS) phase is realized when $\Delta_{+}\Delta_{-}<0$. (b) Josephson junction between two TRITOPSs.\label{fig:setup_and_spectrum}}
\end{figure}

%\paragraph*{Low-energy model.\textemdash{}}
\section{Low-energy model}

A uniform TRITOPS can be described, at low energies, by the Hamiltonian~\citep{Qi2010topological,Haim2016interaction,Haim2019time}
\begin{equation}
\begin{split}H_{0}=\int_{-\infty}^{\infty}{\rm d}x & \Big\{-iv\sum_{s=\uparrow\downarrow}\left[R_{s}^{\dagger}\partial_{x}R_{s}-L_{s}^{\dagger}\partial_{x}L_{s}\right]\\
 & +\left[\Delta_{+}R_{\uparrow}^{\dagger}L_{\downarrow}^{\dagger}+\Delta_{-}L_{\uparrow}^{\dagger}R_{\downarrow}^{\dagger}+{\rm h.c.}\right]\Big\},
\end{split}
\label{eq:H_0_low_E}
\end{equation}
where $R_{s}(x)$ {[}$L_{s}(x)${]} is a field describing a right-
(left-) moving electron with spin $s=\uparrow,\downarrow$ and velocity
$v$. The pairing potential $\Delta_{+}$ describes pairing between
modes of positive helicity ($R_{\uparrow}$ and $L_{\downarrow}$),
while $\Delta_{-}$ describes pairing between modes of negative helicity
($L_{\uparrow}$ and $R_{\downarrow}$) {[}see Fig.~\hyperref[fig:setup_and_spectrum]{\ref{fig:setup_and_spectrum}(a)}{]}.

We are interested in systems obeying time-reversal symmetry, implemented
by
\begin{equation}
R_{s}\to i\sigma_{ss'}^{y}L_{s'}\hspace{1em};\hspace{1em}L_{s}\to i\sigma_{ss'}^{y}R_{s'}\hspace{1em};\hspace{1em}i\to-i,
\end{equation}
where $\{\sigma^{\alpha=x,y,z}\}$ are the Pauli matrices. This constrains
the pairing potentials to be real, $\Delta_{\pm}\in\mathbb{R}$. It
can be shown that $H_{0}$ is in the topological phase, with a pair
of Majorana zero modes at each end, when the topological invariant $\mathcal{Q}=\rm{sgn}(\Delta_{+}\Delta_{-})$ is negative~\citep{Qi2010topological,Haim2019time}, namely when the positive-helicity modes
experience a pairing potential with opposite sign to that of the negative-helicity
modes.

Notice that $H_{0}$ obeys a spin-rotation symmetry, $[H_{0},S_{z}]=0$, where $S_{z}=\int{\rm d}x\sigma_{ss'}^{z}(R_{s}^{\dagger}R_{s'}+L_{s}^{\dagger}L_{s'})$~\footnote{Note that the index $s$ in Eq.~(\ref{eq:H_0_low_E}) does not have to represent spin. More generally, $s=\pm1$ can label two pairs of fields, $R_1,L_1$ and $L_{-1},R_{-1}$, related by time-reversal symmetry, respectively. The emergent spin-rotation symmetry is then given by $\tilde{S}_{z}=\sum_{s=\pm1}s(R_{s}^{\dagger}R_{s}+L_{s}^{\dagger}L_{s})$.}. This is not a fundamental symmetry of the TRITOPS phase. It is rather
an emergent symmetry of its long-wavelength description. As we will see below, the junction generally breaks this symmetry,
which will prove crucial for obtaining the $\pi$-junction phase. 

We now consider a superconductor - normal-metal - superconductor (SNS)
junction with two TRITOPSs. To this end, we promote $\Delta_{\pm}$
to have the following spatial dependence,
\begin{equation}
\Delta_{\pm}(x)=\Delta_{\pm}^{0}\theta(|x|-d/2)e^{i\frac{\phi}{2}{\rm sgn}(x)},\label{eq:Delta_x}
\end{equation}
where $\theta(x)$ is the Heaviside step function, $d$ is the junction's length, and $\phi$ is the phase
difference across the junction {[}see Fig.~\hyperref[fig:setup_and_spectrum]{\ref{fig:setup_and_spectrum}(b)}{]}.
Furthermore, we allow for spin-orbit coupling, as well as backscattering,
inside the junction, such that the Hamiltonian for the entire system
is $H=H_{0}+H_{{\rm so}}+H_{{\rm bs}}$, with

\begin{equation}
\begin{split}H_{{\rm so}}= & U_{{\rm so}}\int_{-d/2}^{d/2}{\rm d}x\left(R_{\uparrow}^{\dagger}R_{\downarrow}-L_{\uparrow}^{\dagger}L_{\downarrow}\right)+{\rm h.c.,}\\
H_{{\rm b}}= & V_{\rm b}'\sum_{s=\uparrow\downarrow}\left[R_{s}^{\dagger}(0)+L_{s}^{\dagger}(0)\right]\left[R_{s}(0)+L_{s}(0)\right].
\end{split}
\label{eq:H_so_and_bs}
\end{equation}
Here, $H_{{\rm so}}$ is a spin-orbit coupling term, responsible for rotation of the spin as the electron traverses the junction, and
$H_{{\rm b}}$ is a delta-potential barrier that controls the transparency
of the junction~\footnote{One can consider other kinds of terms that induces backscattering,
such as for example weak links at $x=\pm d/2$. We have checked numerically
that this does not affect the conclusions of this work.}. Both terms are allowed by time-reversal symmetry and therefore will
generally be present.

\begin{figure}
\begin{centering}
\begin{tabular}{lr}
\hskip -2mm
\includegraphics[clip=true,trim=0mm 0mm 0mm 0mm,height=3.3cm]{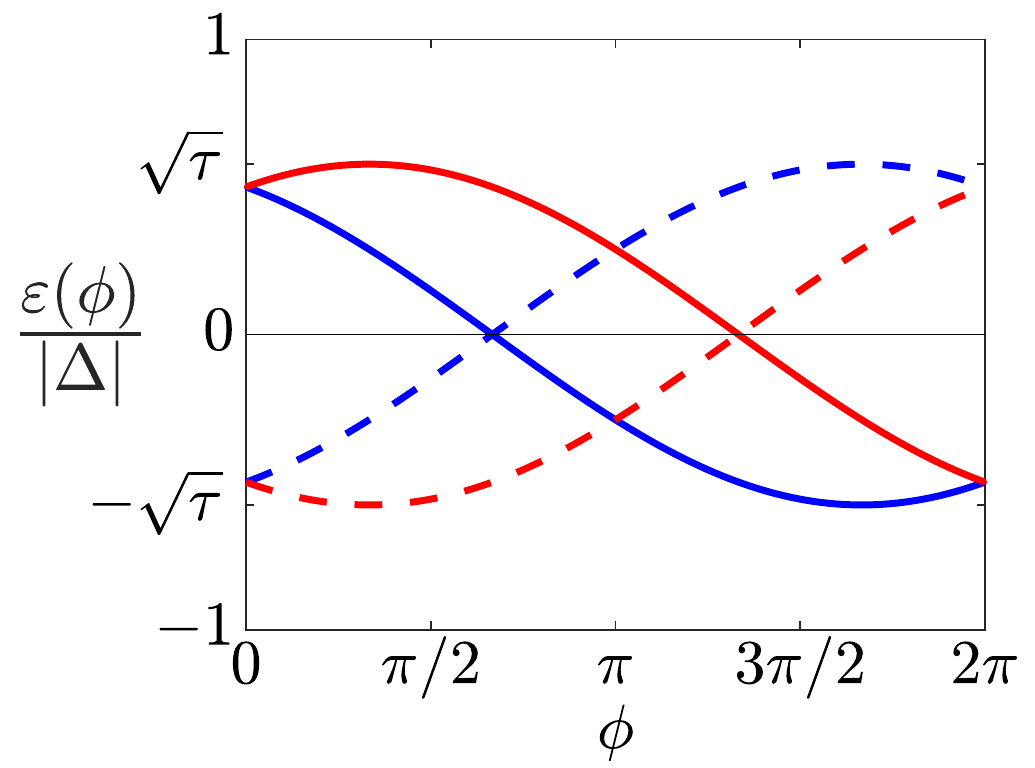}
\llap{\parbox[c]{8.3cm}{\vspace{-2mm}(a)}}
&
\hskip -2mm
\includegraphics[clip=true,trim=0mm 0mm 0mm 0mm,height=3.3cm]{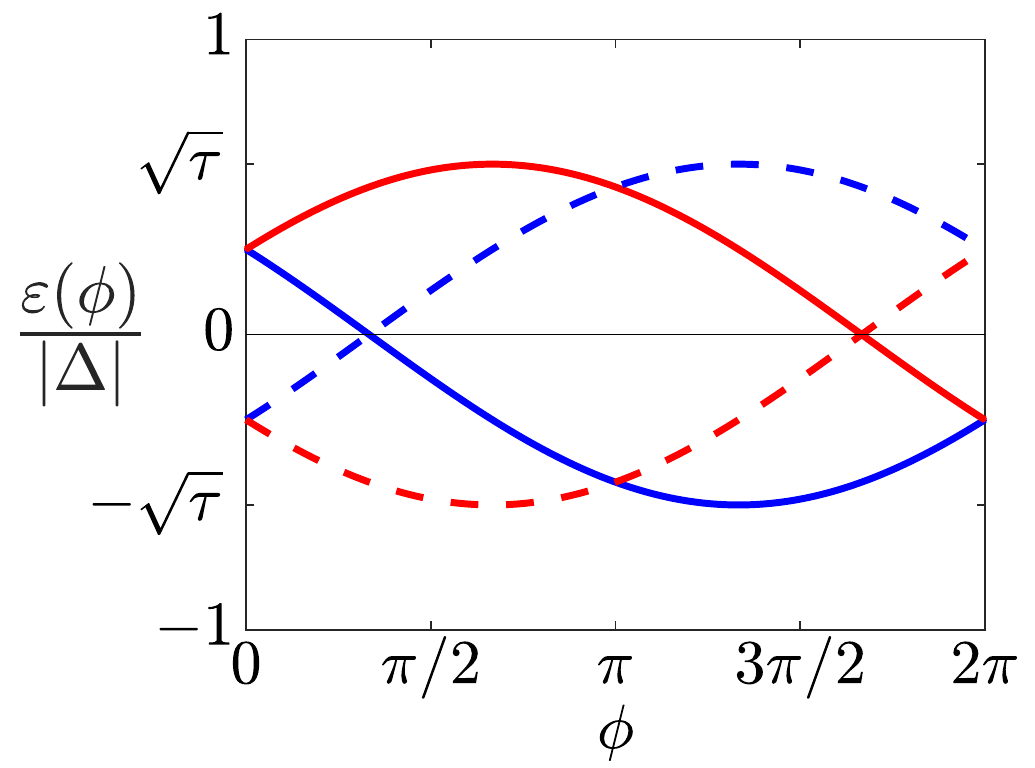}
\llap{\parbox[c]{8.3cm}{\vspace{-2mm}(b)}}
\\
\hskip -2mm
\includegraphics[clip=true,trim=0mm 0mm 0mm 0mm,height=3.3cm]{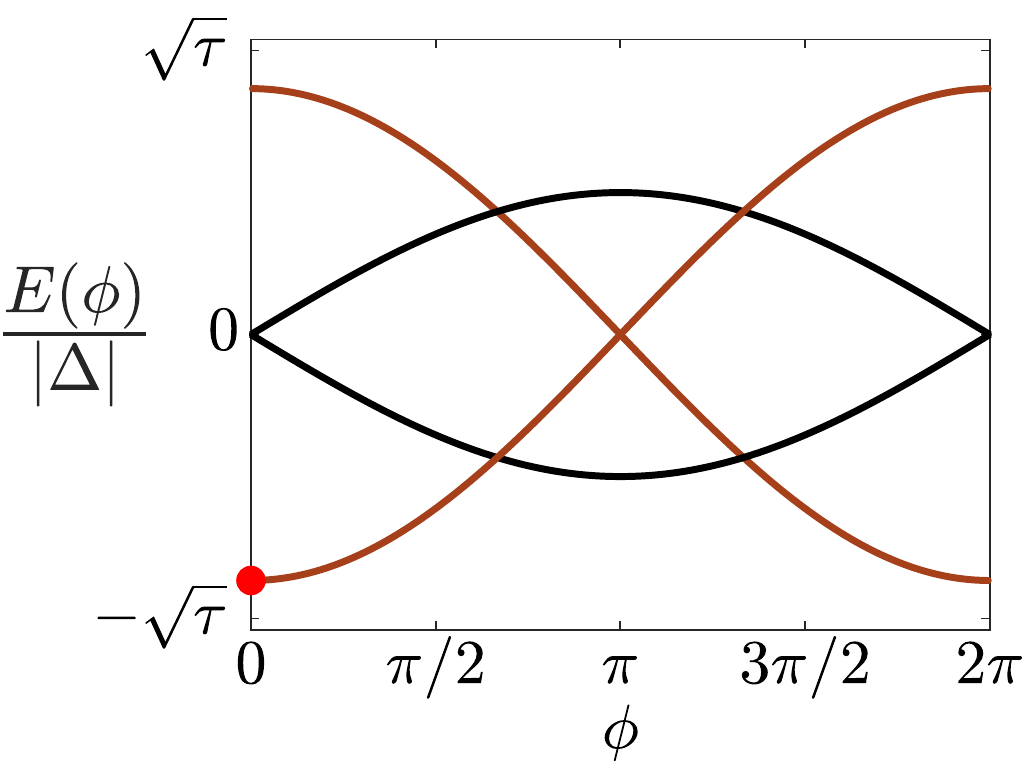}
\llap{\parbox[c]{8.3cm}{\vspace{-2mm}(c)}}
&
\hskip -2mm
\includegraphics[clip=true,trim=0mm 0mm 0mm 0mm,height=3.3cm]{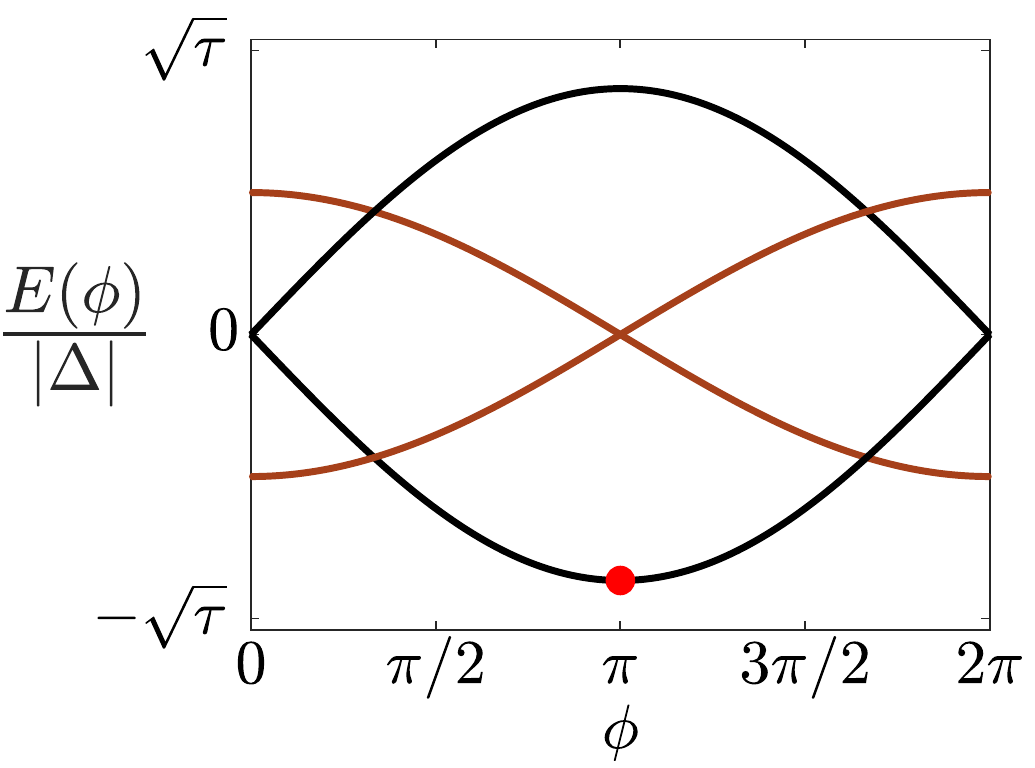}
\llap{\parbox[c]{8.3cm}{\vspace{-2mm}(d)}}
\end{tabular}
\end{centering}

\caption{(a,b) Single-particle excitation spectrum of the Josephson junction
described in Eqs.~(\ref{eq:H_0_low_E},\ref{eq:Delta_x},\ref{eq:H_so_and_bs}), as
a function of phase bias, for different spin-rotation angle: (a) $\beta_{{\rm so}}=\pi/3$,
and (b) $\beta_{{\rm so}}=2\pi/3$. Excitation energies marked by
a solid and dashed line are related by particle-hole symmetry. (c,d)
Many-body energy spectra, corresponding to (a,b), respectively. In
(c), the global ground state is at $\phi_{{\rm eq}}=0$, while in
(d) the global ground state is at $\phi_{{\rm eq}}=\pi$. This transition
from a $0$-junction to a $\pi$-junction occurs abruptly at $\beta_{{\rm so}}=\pi/2$
{[}see Eq.~(\ref{eq:phi_eq}){]}. Energy states marked in brown (black)
have even (odd) Fermion parity. \label{fig:Low_E_spec}}
\end{figure}

To obtain the ground-state energy of the junction, we first calculate
the spectrum of Andreev bound states, which is done by solving the
single-particle Schr\"odinger equation for $H$. In the special case
of $\Delta_{+}^{0}=-\Delta_{-}^{0}\equiv\Delta$, and $d\ll v/\left|\Delta\right|$,
one obtains~\citep{Mellars2016signatures,Kwon2004fractional} 
\begin{equation}
\varepsilon_{\pm}(\phi)=\sqrt{\tau}\Delta\cos\left(\frac{\phi\pm\beta_{{\rm so}}}{2}\right)\,\,\,\,;\,\,\,\,\beta_{{\rm so}}\equiv\frac{2\left|U_{{\rm so}}\right|d}{v},\label{eq:eps_pm_phi}
\end{equation}
where $\tau=1/[1+(V_{\rm b}'/v)^{2}]$ is the transmission probability of
the junction in the normal state, and $\beta_{{\rm so}}$ is the spin-rotation
angle acquired as the electron traverses the junction. Together with
their particle-hole partners, the excitation spectrum is given by
$\left\{ \varepsilon_{+}(\phi),\varepsilon_{-}(\phi),-\varepsilon_{+}(\phi),-\varepsilon_{-}(\phi)\right\} $,
shown in Figs.~\hyperref[fig:Low_E_spec]{\ref{fig:Low_E_spec}(a,b)}, for $\beta_{{\rm so}}=\pi/3$
and $\beta_{{\rm so}}=2\pi/3$, respectively~\footnote{We note that there are four zero-energy Majorana bound states, at
the two outer ends of the entire system. Being far away from the junction,
they do not affect the physics, and we therefore ignore them hereafter.}.

For a fixed phase difference, $\phi$, the ground-state energy is
obtained by summing the negative excitation energies,
\begin{equation}
E_{{\rm gs}}(\phi)=-[\left|\varepsilon_{+}(\phi)\right|+\left|\varepsilon_{-}(\phi)\right|]/2.\label{eq:E_gs_phi}
\end{equation}
If the phase difference is not set externally, it is determined, at
zero temperature, by minimization of $E_{{\rm gs}}(\phi)$. From Eqs.~(\ref{eq:eps_pm_phi}) and~(\ref{eq:E_gs_phi}) one then
obtains the phase difference at thermal equilibrium,
\begin{equation}
\phi_{{\rm eq}}=\begin{cases}
0, & |\tan(\beta_{{\rm so}}/2)|<1\\
\pi, & |\tan(\beta_{{\rm so}}/2)|>1.
\end{cases}\label{eq:phi_eq}
\end{equation}
As $\beta_{{\rm so}}$ is varied, the system goes through a series
of transitions between a $0$-junction and a $\pi$-junction, at $\beta_{{\rm so}}=\pi(1/2+N)$,
for integer $N$. This can be achieved by tuning the length of the
junction, $d$, or the velocity, $v$, which generally depends on
the chemical potential. Figures~\hyperref[fig:Low_E_spec]{\ref{fig:Low_E_spec}(c,d)} present
the four lowest many-body energies, obtained by summing the single-particle
excitation energies according to their occupation, for $\beta_{{\rm so}}=\pi/3$
and $\beta_{{\rm so}}=2\pi/3$. In the former case the minimal energy
is obtained for $\phi_{{\rm eq}}=0$, while in the latter it is obtained
for $\phi_{{\rm eq}}=\pi$.

In the case of finite temperature, $T,$ the equilibrium phase is
determined by minimizing the free energy, $F(\phi)=-T\sum_{p=\pm}\ln\{2\cosh[\varepsilon_p(\phi)/2T]\}$,
instead of the ground-state energy. Within the limits of validity of
Eq.~(\ref{eq:eps_pm_phi}), one can check that this does not affect
the result for $\phi_{{\rm eq}}$, Eq.~(\ref{eq:phi_eq})~\footnote{Note that, while high-excited states contribute to $F$ at finite temperature, their dependence on $\phi$ is negligible.}.

Finally, while the spectrum in Eq.~(\ref{eq:eps_pm_phi}) was calculated
under the simplifying assumptions, $\Delta_{+}=-\Delta_{-}=\Delta$
and $d\ll v/|\Delta|$, its qualitative features are universal~\citep{Zhang2014anomalous,Mellars2016signatures}.
Specifically, the level crossings at $\phi=0,\pi$ are protected by time-reversal
symmetry, and the crossings at zero energy are protected by particle-hole
symmetry~\footnote{In the many-body spectrum {[}see Figs.~\hyperref[fig:Low_E_spec]{\ref{fig:Low_E_spec}(c,d)}{]},
this is manifested in the protection of crossings between states of
even- and odd-Fermion parity, including against non-quadratic
perturbations.}.

%\paragraph*{Physical picture.\textemdash{}}
\section{Physical picture}

\begin{figure}
\begin{centering}
%\vspace{2mm}
\begin{tabular}{lr}
\hskip -2mm
\includegraphics[clip=true,trim=0mm 0mm 0mm 0mm,height=1.9cm]{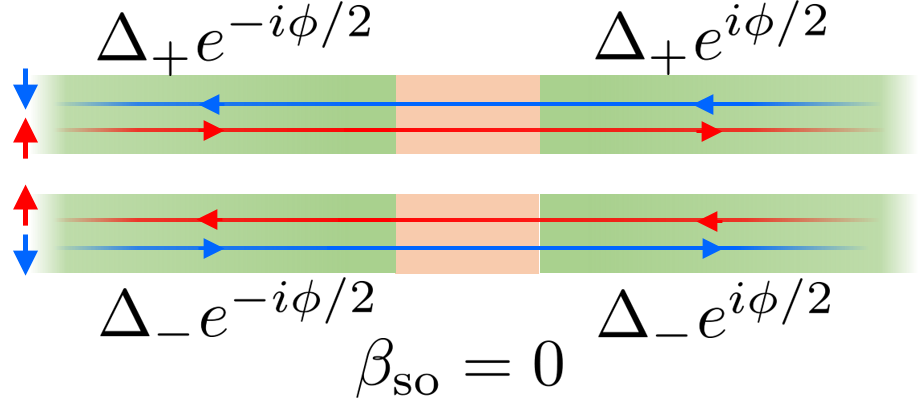}
\llap{\parbox[c]{8.4cm}{\vspace{0.1cm}(a)}}
&
\includegraphics[clip=true,trim=0mm 0mm 0mm 0mm,height=1.9cm]{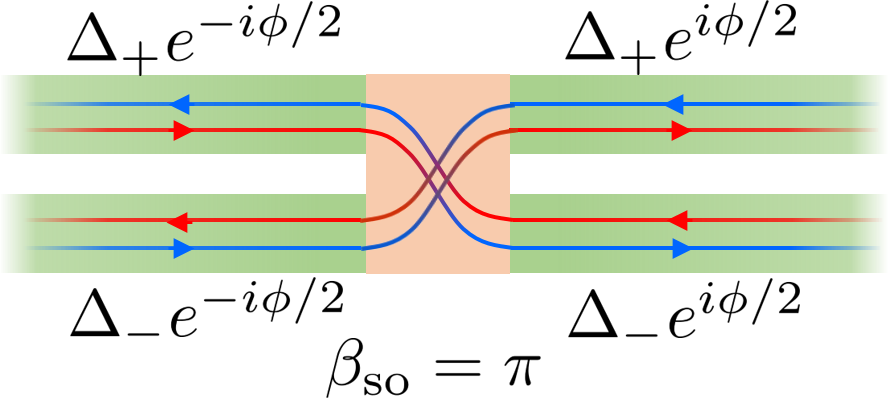}
\llap{\parbox[c]{8.4cm}{\vspace{0.1cm}(b)}}
\end{tabular}
%\vskip -1mm
\end{centering}

\caption{(a) In the absence of spin rotation in the junction, modes of positive
and negative helicity do not mix. The Josephson coupling therefore
tends to aligns the phases of $\Delta_{\pm}(x<0)$ with those of $\Delta_{\pm}(x>0)$,
respectively, resulting in $\phi_{{\rm eq}}=0$. (b) At the other
extreme, when the spin-rotation angle is $\beta_{{\rm so}}=\pi$,
modes of positive (negative) helicity are converted to modes of negative
(positive) helicity. Josephson coupling then tends to align the phases
of $\Delta_{\pm}(x<0)$ with those of $\Delta_{\mp}(x>0)$, respectively.
In the TRITOPS phase, where $\Delta_{+}\Delta_{-}<0$, this means
$\phi_{{\rm eq}}=\pi.$\label{fig:phys_pic}}
\end{figure}

The above results can be intuitively understood from the low-energy
description of the TRITOPS phase, Eq.~(\ref{eq:H_0_low_E}). In the
absence of spin-orbit coupling in the junction ($\beta_{{\rm so}}=0$),
the Josephson junction decouples into two separate junctions: one
involving the positive-helicity modes (with pairing potential $\Delta_{+}$),
and one involving the negative-helicity modes (with pairing potential
$\Delta_{-}$), as depicted in Fig.~\hyperref[fig:phys_pic]{\ref{fig:phys_pic}(a)}. The Josephson
coupling across the junction then seeks to align the phases of $\Delta_{\pm}(x<0)$
with the phases of $\Delta_{\pm}(x>0)$, respectively, which is achieved when $\phi=0$.

For a non-vanishing spin-orbit coupling in the junction, on the other
hand, the electron's spin rotates by an angle $\beta_{{\rm so}}$
as it traverses the junction, causing modes of positive and negative
helicities to mix. In the special case of $\beta_{{\rm so}}=\pi$,
modes of positive helicity are perfectly converted to mode of negative
helicity and vice versa, as depicted in Fig.~\hyperref[fig:phys_pic]{\ref{fig:phys_pic}(b)}.
The Josephson coupling now seeks to align the phases of $\Delta_{\pm}(x<0)$
with those of $\Delta_{\mp}(x>0)$, respectively. Importantly, since in the TRITOPS phase $\Delta_{+}\Delta_{-}<0$, this translates to having $\phi=\pi$.
The transition between these two cases occurs at
$\beta_{{\rm so}}=\pi/2$ {[}See Eq.~(\ref{eq:phi_eq}){]}.

%\paragraph*{Experimental signature.\textemdash{}}
\section{Experimental signature}

The transition between a $0$-junction and a $\pi$-junction induces
an abrupt change in the system's physical observables, as in a first-order
phase transition. Below, we focus on the behavior of the tunneling
density of states and the critical current,
and propose these can serve as experimental signatures of the transition.

We wish to study the Josephson junction beyond the simplifying assumptions
leading to Eq.~(\ref{eq:eps_pm_phi}). To this end, we consider a
lattice model of a TRITOPS and analyze it numerically. For a uniform
TRITOPS, the Hamiltonian is given by~\citep{Zhang2013time} 
\begin{equation}
\begin{split}H_{0}^{{\rm Latt}}= & \sum_{n}\bigg\{-\mu\boldsymbol{\boldsymbol{c}}_{n}^{\dagger}\boldsymbol{\boldsymbol{c}}_{n}-\left[\boldsymbol{\boldsymbol{c}}_{n}^{\dagger}\left(t+iu\sigma_{z}\right)\boldsymbol{\boldsymbol{c}}_{n+1}+{\rm h.c.}\right]\\
 & +\left[\frac{1}{2}\Delta_{0}\boldsymbol{\boldsymbol{c}}_{n}^{\dagger}i\sigma_{y}\boldsymbol{\boldsymbol{c}}_{n}^{\dagger{\rm T}}+\Delta_{1}\boldsymbol{\boldsymbol{c}}_{n}^{\dagger}i\sigma_{y}\boldsymbol{\boldsymbol{c}}_{n+1}^{\dagger{\rm T}}+{\rm h.c.}\right]\bigg\},
\end{split}
\label{eq:H_0_latt}
\end{equation}
where $\boldsymbol{\boldsymbol{c}}_{n}^{\dagger}=(c_{n\uparrow}^{\dagger},c_{n\downarrow}^{\dagger})$,
and $c_{ns}^{\dagger}$ creates an electron on site $n$ with spin
$s$. Here, $\mu$ is the chemical potential, $t$ is a hopping parameter,
$u$ is the spin-orbit coupling coefficient, and $\Delta_{{\rm 0}}$
and $\Delta_{1}$ are singlet pairing potentials describing on-site
and nearest-neighbor pairing, respectively. The system is in the
topological phase when~\citep{Zhang2013time} $2\left|u\right|\sqrt{1-\left[\Delta_{0}/(2\Delta_{1})\right]^{2}}>|\mu-t\Delta_{0}/\Delta_{1}|$.

Before proceeding, it is instructive to relate the lattice Hamiltonian,
Eq.~(\ref{eq:H_0_latt}), to the low-energy Hamiltonian of Eq.~(\ref{eq:H_0_low_E}).
This can be done, in the weak pairing limit~\footnote{In this limit the pairing potentials are small compared with the Fermi
level, measured from the bottom of the band, $\left|\Delta_{{\rm 0}}\right|,\left|\Delta_{1}\right|\ll2\sqrt{t^{2}+u^{2}}-|\mu|$.}, by linearizing the spectrum of $H_{0}^{{\rm Latt}}$ near the Fermi
momenta, $a_{0}k_{{\rm F},\pm}=\mp\lambda+\cos^{-1}\left[-\mu/(2w)\right]$,
where $t=w\cos(\lambda)$, $u=w\sin(\lambda)$, and $a_{0}$ is the
lattice constant {[}see also Fig.~\hyperref[fig:setup_and_spectrum]{\ref{fig:setup_and_spectrum}(a)}{]}.
The pairing potentials in the linearized model, Eq.~(\ref{eq:H_0_low_E}),
are then given by $\Delta_{\pm}=\Delta_{0}+2\Delta_{1}\cos(a_{0}k_{{\rm F},\pm})$,
and the velocity is $v=a_{0}\sqrt{4w^{2}-\mu^{2}}.$

As before, in order to simulate a Josephson junction, we take the pairing
potentials to depend on position according to $\Delta_{0,1}(n)=\Delta_{0,1}\theta(|na_{0}|-d/2)\exp\left[i{\rm sgn}(n)\phi/2\right]$.
To account for spin rotation inside the junction, we include a spin-orbit
coupling term in a perpendicular direction to the one in the bulk.
This is done by letting $u$, in Eq.~(\ref{eq:H_0_latt}), vanish inside the junction, and instead adding a term~\footnote{Similar results are obtained for spin-orbit coupling in the $y$ direction.},
\begin{equation}
\begin{split}H_{{\rm so}}^{{\rm Latt}}=iu_{{\rm so},{\rm J}}\sum_{\left|na\right|<d/2} & \left[\boldsymbol{\boldsymbol{c}}_{n}^{\dagger}\sigma_{x}\boldsymbol{\boldsymbol{c}}_{n+1}-{\rm h.c.}\right]\end{split}
.\label{eq:H_so_latt}
\end{equation}
For small $u_{{\rm so,J}},\mu$, the resulting spin rotation angle is $\beta_{{\rm so}}\simeq u_{{\rm so,J}}d/(a_{0}t)$.
Finally, backscattering is accounted for by $H_{{\rm b}}^{{\rm Latt}}=V_{{\rm b}}\boldsymbol{c}_{0}^{\dagger}\boldsymbol{c}_{0}$,
such that altogether the Hamiltonian is given by $H^{{\rm Latt}}=H_{0}^{{\rm Latt}}+H_{{\rm so}}^{{\rm Latt}}+H_{{\rm b}}^{{\rm Latt}}.$

\begin{figure}
\begin{centering}
\vspace{-2.5mm}
\begin{tabular}{lr}
\hskip -2mm
\includegraphics[clip=true,trim=0mm 0mm 0mm 0mm,height=3.35cm]{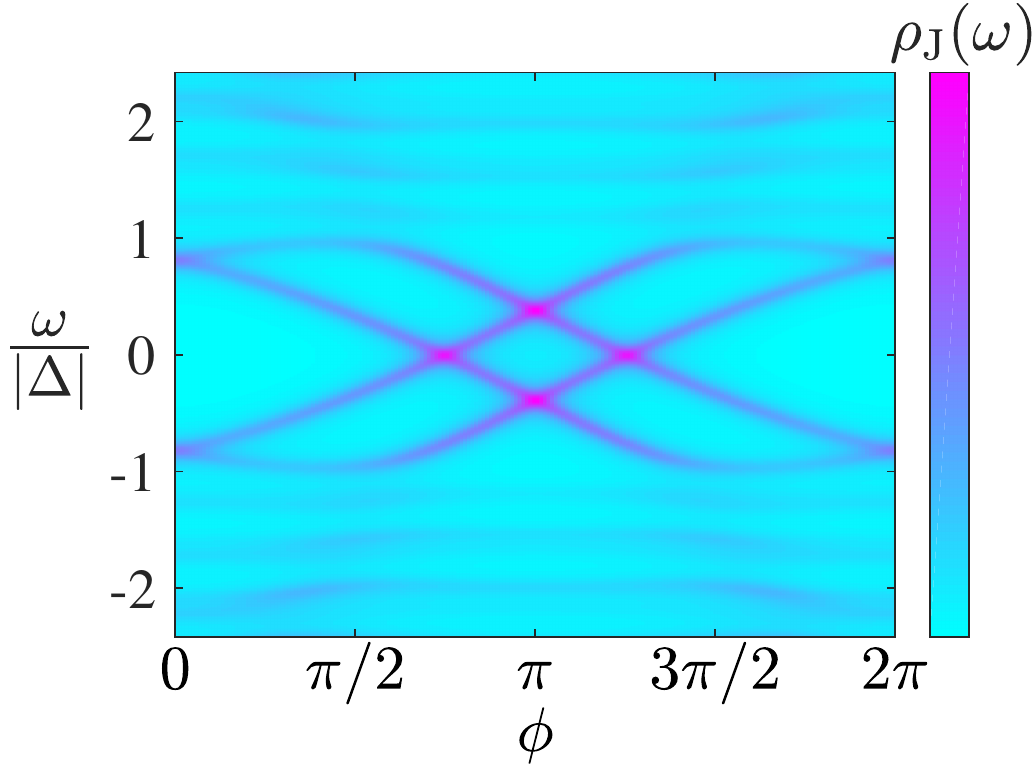}
\llap{\parbox[c]{8.5cm}{\vspace{-2mm}(a)}}
&
\hskip -2.5mm
\includegraphics[clip=true,trim=0mm 0mm 0mm 0mm,height=3.35cm]{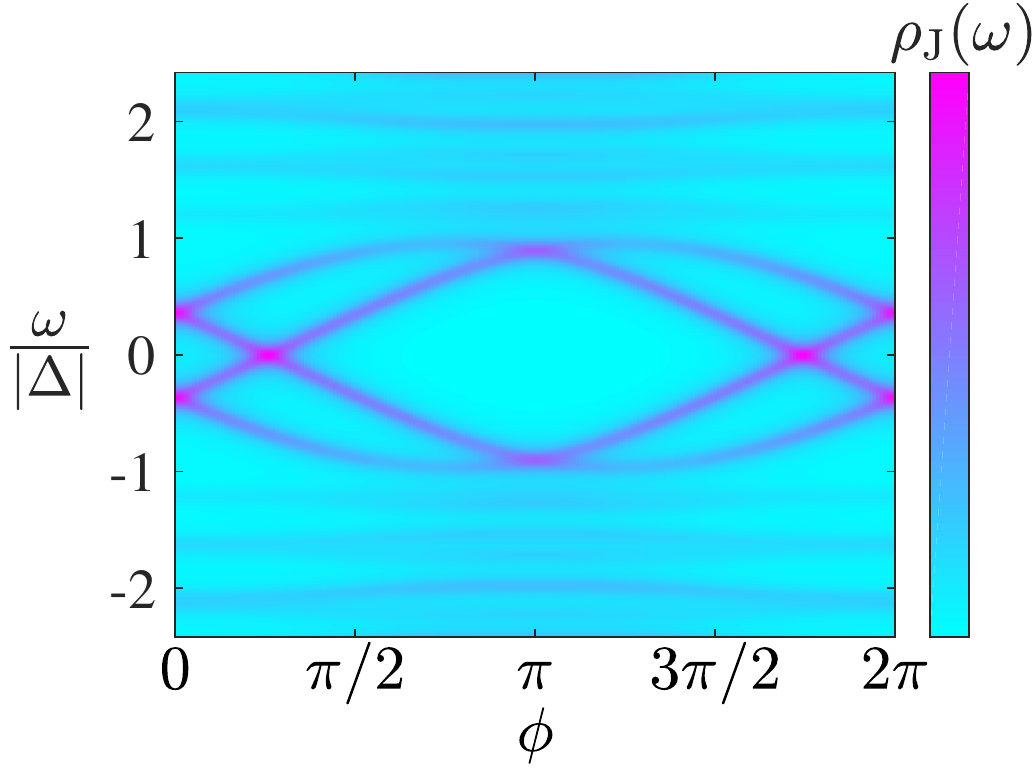}
\llap{\parbox[c]{8.5cm}{\vspace{-2mm}(b)}}
\\
\hskip -2mm
\includegraphics[clip=true,trim=0mm 0mm 0mm 0mm,height=3.35cm]{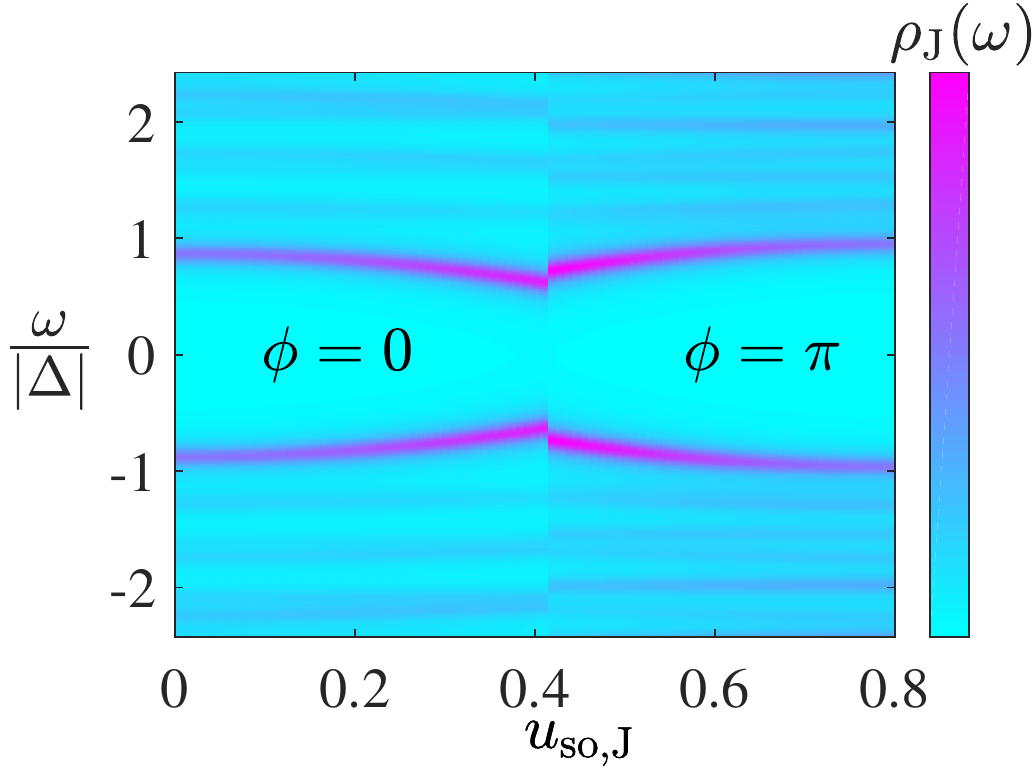}
\llap{\parbox[c]{8.5cm}{\vspace{-2mm}(c)}}
&
\hskip -2.5mm
\includegraphics[clip=true,trim=0mm 0mm 0mm 0mm,height=3.35cm]{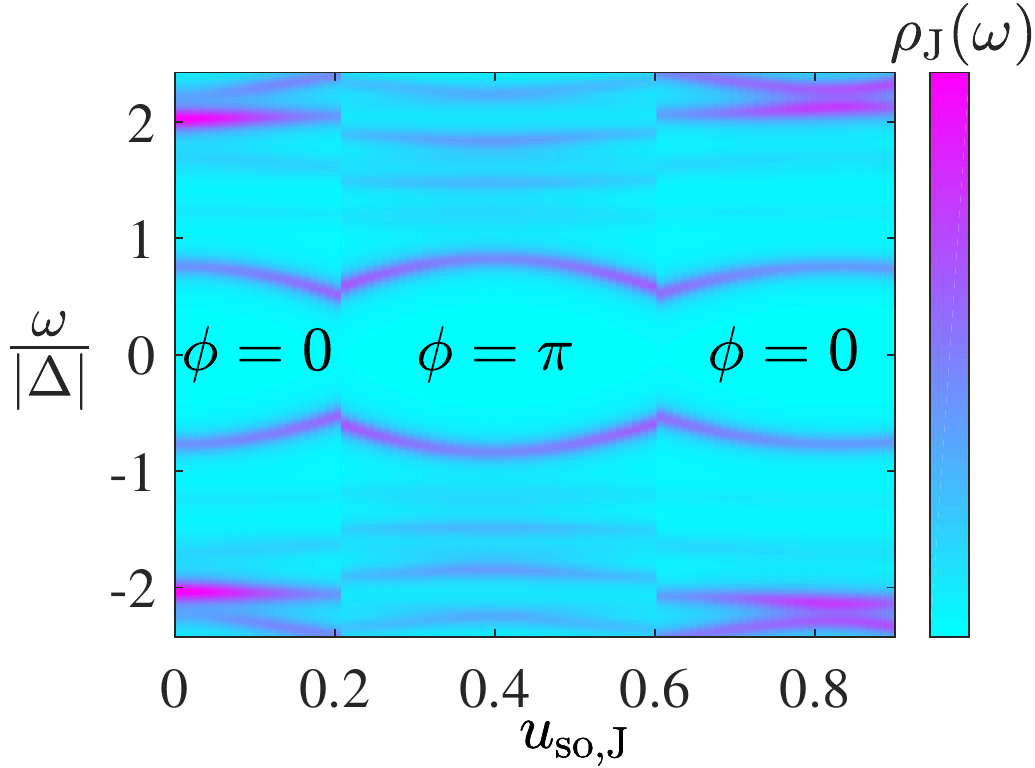}
\llap{\parbox[c]{8.5cm}{\vspace{-2mm}(d)}}
\end{tabular}
\end{centering}

\caption{Density of states inside the junction, $\rho_{{\rm J}}(\omega)$,
in arbitrary units, for fixed parameters, $t=2$, $u=1$, $\mu=-1.5$,
$\Delta_{{\rm 0}}=0.2$, $\Delta_{{\rm 1}}=1$, and $V_{{\rm b}}=1$
{[}see Eqs.~(\ref{eq:H_0_latt},\ref{eq:H_so_latt}){]}. Each of
the superconductors is of length $L_{{\rm sc}}=100a_{0}$. In (a,b),
$\rho_{{\rm J}}(\omega)$ is presented versus the phase
difference, $\phi$, for junction length $d=5a_{0}$, and for (a)
$u_{{\rm so,J}}=0.2$, (b) $u_{{\rm so,J}}=0.6$. In (c,d), $\rho_{{\rm J}}(\omega)$
is presented versus $u_{{\rm so,J}}$, for (c) $d=5a_{0}$,
and (d) $d=9a_{0}$. The transitions between the $0$-junction and
the $\pi$-junction are manifested in a non-analytic behavior of $\rho_{{\rm J}}(\omega)$.
The frequency, $\omega$, is normalized by the bulk gap, $|\Delta|=0.248$.
\label{fig:LDOS}}
\end{figure}

To analyze $H^{{\rm Latt}}$, we first rewrite it in a Bogoliubov-de
Gennes form, $H^{{\rm Latt}}=\frac{1}{2}\sum_{nn'}\psi_{n}^{\dagger}\mathcal{H}_{nn'}\psi_{n'}$,
where $\psi_{n}^{\dagger}=(\boldsymbol{c}_{n}^{\dagger},\boldsymbol{c}_{n}^{\rm T})$,
and accordingly $\mathcal{H}_{nn'}$ is a $4\times4$ matrix that
includes spin and particle-hole degrees of freedom. The tunneling
density of states inside the junction is then given by
\begin{equation}
\rho_{{\rm J}}(\omega)=-\frac{1}{\pi}{\rm Im}\sum_{|na_{0}|<d/2}{\rm Tr}G_{nn}^{R}(\omega),
\end{equation}
where $G^{R}(\omega)=\left[\omega+i\eta-\mathcal{H}\right]^{-1}$ is the retarded Green function~\footnote{In the following simulations we use $\eta=0.02$. Experimentally, this parameter is determined by the coupling of the junction to the probe.}. As a preliminary, we
calculate the tunneling density of states as a function of the phase
difference, $\phi$, for fixed system parameters in the TRITOPS phase.
The results are presented in Figs.~\hyperref[fig:LDOS]{\ref{fig:LDOS}(a,b)} for $u_{{\rm so,J}}=0.2$
and $u_{{\rm so,J}}=0.6$, respectively, with $d=5a_{0}$, and $V_{{\rm b}}=1$.
This should be compared with the excitation spectrum given in Eq.~(\ref{eq:eps_pm_phi})
and shown in Figs.~\hyperref[fig:Low_E_spec]{\ref{fig:Low_E_spec}(a,b)}. Notice that while
the latter was obtained from the linearized model in the limit of
a short junction and for the special case $\Delta_{+}=-\Delta_{-}$,
the qualitative features of the spectrum are retained.

Next, we wish to examine the transition between a 0-junction and a $\pi$-junction
when tuning one of the system parameters. We vary the spin-orbit-coupling
coefficient in the junction, $u_{{\rm so,J}}$. For each value of
$u_{{\rm so,J}}$, we numerically search for the phase, $\phi_{{\rm eq}}$
, which minimizes the ground state energy of $H^{{\rm Latt}}$,

\begin{equation}
E_{{\rm gs}}(\phi)=\frac{1}{2}\sum_{\varepsilon_{\nu}<0}\varepsilon_{\nu}(\phi),
\end{equation}
where $\left\{ \varepsilon_{\nu}(\phi)\right\} _{\nu}$ are the eigenvalues
of $\mathcal{H}$. Figures~\hyperref[fig:LDOS]{\ref{fig:LDOS}(c,d)} present $\rho_{{\rm J}}(\omega;\phi_{{\rm eq}})$
for two different junction's length, $d=5a_{0}$ and $d=9a_{0}$,
respectively. As expected, the density of states exhibits non-analytic
behavior at the transitions. This should be compared with Eqs.~(\ref{eq:eps_pm_phi},\ref{eq:phi_eq}),
which suggest that, at the transition, the subgap excitations,
$\varepsilon_{\pm}=\sqrt{\tau/2}\Delta$, are continuous, but have
a jump in their derivative.

\begin{figure}
\begin{centering}
\begin{tabular}{lr}
\hskip -1mm
\includegraphics[clip=true,trim=0mm 0mm 0mm 0mm,height=3.3cm]{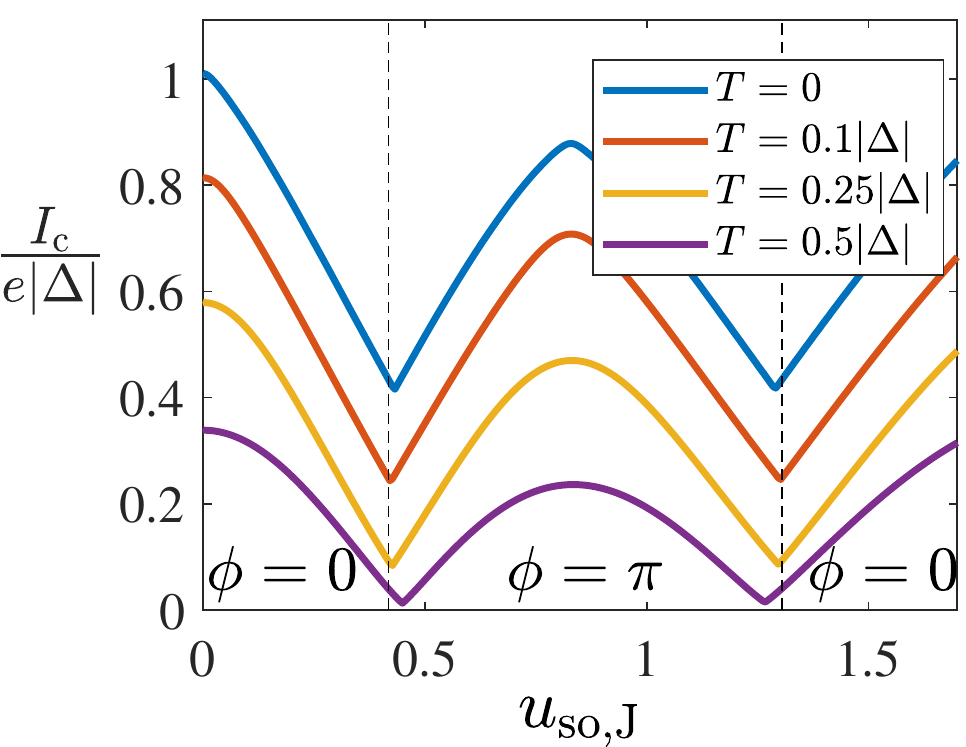}
\llap{\parbox[c]{8cm}{\vspace{0mm}(a)}}
&
\hskip -1mm
\includegraphics[clip=true,trim=0mm 0mm 0mm 0mm,height=3.3cm]{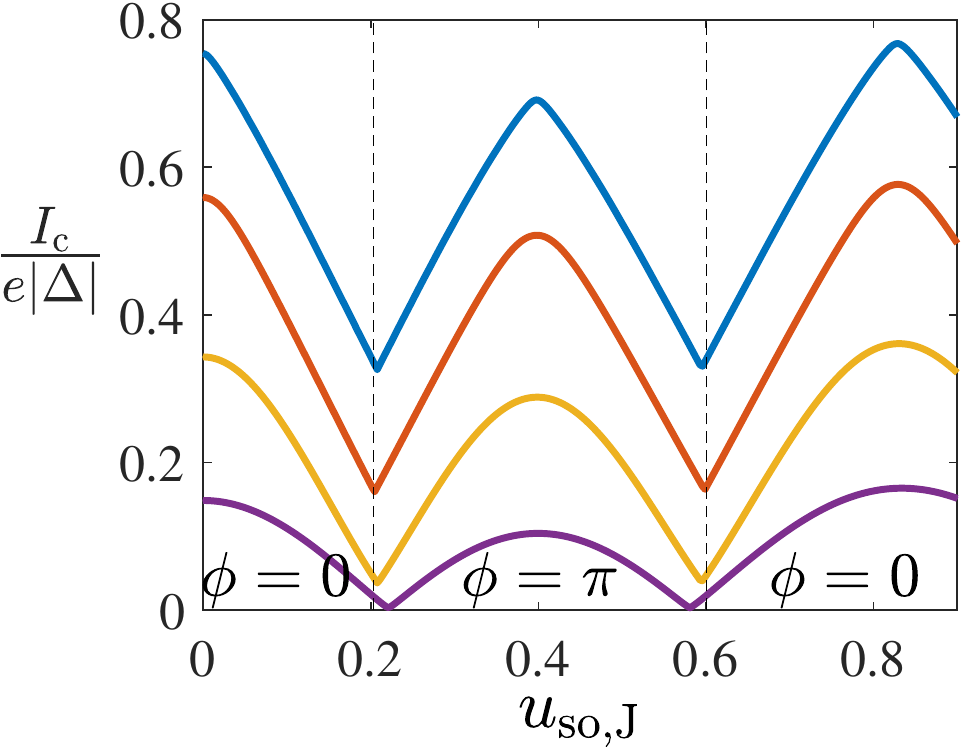}
\llap{\parbox[c]{8cm}{\vspace{0mm}(b)}}
\end{tabular}
%\vskip -1mm
\end{centering}

\caption{(a,b) Critical current as a function of the spin-orbit coupling in
the junction, for different temperatures. System parameters are the
same as in Figs.~\hyperref[fig:LDOS]{\ref{fig:LDOS}(c,d)}, respectively. Vertical dashed
lines mark the zero-temperature transitions between a 0 junction and
a $\pi$ junction. At these points, $I_{{\rm c}}$ has a minimum and
a discontinuity in its derivative.\label{fig:crit_current}}
\end{figure}

A signature of the transition can also be found in measurement of
the critical current, given by $I_{{\rm c}}={\rm max}_{\phi}[|I_{{\rm s}}(\phi)|]$,
where $I_{{\rm s}}(\phi)=2e{\rm d}F(\phi)/{\rm d}\phi$ is the supercurrent
for fixed $\phi$. Within the limits of validity of Eqs.~(\ref{eq:eps_pm_phi},\ref{eq:E_gs_phi}),
and for zero temperature, one arrives at $I_{c}=e\sqrt{\tau}|\Delta|\max\left\{ \cos^{2}(\beta_{{\rm so}}/2),\sin^{2}(\beta_{{\rm so}}/2)\right\} $.
Importantly, at the transition points, $\beta_{{\rm so}}=\pi(1/2+N)$,
the critical current has its minimum, accompanied by a discontinuity
in ${\rm d}I_{{\rm c}}/{\rm d}\beta_{{\rm so}}$. In Figs.~\hyperref[fig:crit_current]{\ref{fig:crit_current}(a,b)}
we present $I_{{\rm c}}$ versus the junction's spin-orbit coupling,
for different temperatures, calculated from the lattice model, Eqs.~(\ref{eq:H_0_latt},\ref{eq:H_so_latt}),
for the same parameters as in Figs.~\hyperref[fig:LDOS]{\ref{fig:LDOS}(c,d)}, respectively.
Indeed, $I_{{\rm c}}$ has a non-analytic minimum at the transitions.

%\paragraph*{Discussion.\textemdash{}}
\section{Discussion}

We have shown that a $\pi$ junction can spontaneously form, in the absence of magnetic fields, between two topological superconductors. Unlike its topologically-trivial counter part, this Josephson $\pi$ junction does not form as a result of Coulomb blockade. Instead, it is driven by spin-orbit coupling in the junction. When varying the rotation angle acquired by the electron's spin as it passes the junction, the system goes through multiple transitions between a $0$-junction, where
the phase difference at thermal equilibrium is $\phi_{{\rm eq}}=0$,
and a $\pi$-junction, where the phase difference is $\phi_{{\rm eq}}=\pi$. 

Experimentally, these transitions should be observable when one avoids
fixing the phase externally (for instance using a flux loop), but
rather let the phase be determined based on energetic considerations.
In particular, if the TRITOPSs in the junction are realized by a semiconductor-superconductor
heterostructure~\citep{Nakosai2013majorana,Wong2012majorana,Zhang2013time,Keselman2013inducing,Gaidamauskas2014majorana,Haim2014time,Klinovaja2014Kramers,Klinovaja2014time,Schrade2015proximity,Danon2015interaction,Haim2016interaction,Thakurathi2018majorana,Ebisu2016theory,Hsu2018majorana,Wang2018high,Yan2018majorana,Baba2018Cooper},
it is important to avoid direct coupling between the parent superconductors,
as this can give rise to a Josephson coupling that competes with the
mechanism studied above.

We propose measuring the density of states in the junction as a way
of observing the transitions. This can be done, for example, using
a weakly-coupled metallic lead or an STM probe. The density of states
exhibits non-analytic behavior at the transitions {[}see Figs.~\hyperref[fig:LDOS]{\ref{fig:LDOS}(a,b)}{]}.
To tune across the transition, one has to vary parameters which control
the spin rotation angle in the junction, such as the junction's length
or the electron velocity.

As an alternative signature, one can force current through the junction
and measure the critical current. As one tunes across a transition
point, the critical current exhibits a sharp dip, accompanied by a discontinuity
in its first derivative. Similar behavior has been recently predicted~\citep{Pientka2017topological}
in a different context, in a planar Josephson junction. There, a magnetic
field drives a first-order phase transition, where the phase difference
changes discontinuously. Note that, in their case, the phase on either
side of the transition is not limited to the values $\{0,\pi\}$.

For a Josephson junction realized in a semiconductor-superconductor
heterostructures, one can estimate the junction's length needed
to observe a transition. Assuming a spin-orbit coupling scale of $U_{{\rm so}}\sim0.1{\rm meV}$,
and electron velocity of $v\sim10^{5}m/s$, the first transition occurs
at a length $d\sim500{\rm nm}$ {[}see Eqs.~(\ref{eq:eps_pm_phi},\ref{eq:phi_eq}){]}.

A particularly appealing system for demonstrating the effect is a 2d topological insulator where each edge is coupled to a conventional superconductors~\cite{Keselman2013inducing,Klinovaja2014Kramers}, thereby realizing a similar scenario to the one depicted in Fig.~\ref{fig:phys_pic}. For each side of the junction to be in the TRITOPS phase, the superconductors on the upper and lower edges are \emph{tuned} to have a relative $\pi$ phase difference. This, however, does not yet determine the relative phase between the superconductors on the left and right. The prediction of this paper is that the latter will go through a series of transitions between $0$ and $\pi$ as a function of the spin-rotation angle in the junction.

%\paragraph*{Acknowledgments.\textemdash{}}
\section*{Acknowledgments}

I benefited from discussions with L. Arrachea, F. von Oppen, Y. Oreg,
and A. Stern. This research was supported by the Walter Burke Institute
for Theoretical Physics at Caltech.

\bibliographystyle{apsrev4-1}
\bibliography{References}

\end{document}